\documentclass[12pt]{article}

\usepackage{todonotes}

\usepackage{amsmath,amssymb,amsbsy,amsfonts,amsthm,latexsym,
                        amsopn,amstext,amsxtra,euscript,amscd,color}
\usepackage{color,xcolor}
\usepackage{latexsym}
\usepackage{cite}
\usepackage{amsfonts}
\usepackage{amsfonts,mathrsfs}
\usepackage{graphicx}
\usepackage{psfrag}
\usepackage{subfigure}
\usepackage{url}
\usepackage{stfloats}
\usepackage{amsmath}
\usepackage{algorithm}
\usepackage{algorithmic}
\usepackage{hyperref}

\newcommand{\aw}[1]{{\color{black} #1}}

\hypersetup{colorlinks=true}

\newtheorem{theorem}{Theorem}
\newtheorem{lemma}{Lemma}
\newtheorem{corollary}{Corollary}

\newcommand{\quash}[1]{}

\begin{document}

\title{
\aw{Maximum-order complexity and $2$-adic complexity}
}

\date{}

\author{Zhiru~Chen$^1$, Zhixiong~Chen$^2$, Jakob~Obrovsky$^3$ and Arne~Winterhof$^3$\\
~\\
$^1$ State Key Laboratory of Information Security, \\
Institute of Information Engineering,\\
 Chinese Academy of Sciences, Beijing, China\\
$^2$  Fujian Key Laboratory of Financial Information Processing,\\
 Putian University, Putian, Fujian 351100, P. R. China\\
$^3$ Johann Radon Institute for Computational and Applied Mathematics,\\
Austrian Academy of Sciences,
Linz, Austria
}

\maketitle

\begin{center} \aw{Dedicated to the memory of Kai-Uwe Schmidt (1978-2023)}
\end{center}

\begin{abstract}
The $2$-adic complexity has been well-analyzed in the periodic case. However, we are not aware of any theoretical results on the $N$th $2$-adic complexity of any promising candidate for a pseudorandom sequence of finite length $N$ or results on a part of the \aw{period} of length $N$ of a periodic sequence, respectively. Here we introduce the first method for this aperiodic case. More precisely, we study the relation between $N$th maximum-order complexity and $N$th $2$-adic complexity of binary sequences and prove a lower bound on the $N$th $2$-adic complexity in terms of the $N$th maximum-order complexity.
Then any known lower bound on the $N$th maximum-order complexity implies a lower bound on the $N$th $2$-adic complexity of the same order of magnitude. 
In the periodic case, 
one can prove a slightly better result.
The latter bound is sharp which is illustrated by the maximum-order complexity of $\ell$-sequences.
The idea of the proof helps us to characterize the maximum-order complexity of periodic sequences in terms of the unique rational number defined by the sequence.
We also show that a periodic sequence of maximal maximum-order complexity must be also of maximal $2$-adic complexity.
\end{abstract}



\section{Introduction}

\aw{Pseudorandom sequences are generated by deterministic algorithms and are not random at all. However, both from an academic point of view as well as from a cryptographic point of view they should have as many desirable features of randomness as possible, that is, they should not be distinguishable from a 'truly' random sequence. These desirable features and thus
the} pseudorandomness of {\em binary} sequences can be analyzed via several measures of pseudorandomness such as the maximum-order complexity and the $2$-adic complexity, see for example the recent survey \cite{W2023}.
We recall the definitions of the $N$th maximum-order complexity and the $N$th $2$-adic complexity.

For a positive integer $N$, the \emph{$N$th maximum-order complexity} (or \emph{$N$th nonlinear complexity}) $M(\mathcal{S},N)$ of
 a binary sequence $\mathcal{S}=(s_n)_{n\geq 0}$ over the two-element field $\mathbb{F}_2=\{0,1\}$ is defined as the smallest positive integer $\aw{m}$
such that there is a polynomial $f(X_1,\ldots,X_m)\in \mathbb{F}_2[X_1,\ldots,X_m]$
with
$$
s_{i+m}=f(s_i,s_{i+1},\ldots,s_{i+m-1})\quad \mbox{ for }0 \le i \le N-m-1,
$$
see~\cite{CGGT2022,ja91,ja89,jabo,LKK2007,nixi14,PM2006,PM2008,R2005,R2006,SZLH2017,XZLJ2019}.
We set $M(\mathcal{S},N)=0$ if $s_0=s_{1}=\ldots=s_{N-2}=s_{N-1}$, that is $s_i=s_0$ for $0\le i\le N-1$, and 
$M(\mathcal{S},N)=N-1$ if $s_0=s_{1}=\ldots=s_{N-2}\neq s_{N-1}$, that is $s_{i+N-1}=s_i+1$, $i=0$. 

The sequence $(M(\mathcal{S},N))_{N\geq 1}$
is referred to as the \emph{maximum-order complexity profile} (or \emph{nonlinear complexity profile}) of $\mathcal{S}$.
If $\mathcal{S}$ is $T$-periodic, \aw{that is, $s_{n+T}=s_n$ for $n\ge 0$,} we have 
$M(\mathcal{S},N)=M(\mathcal{S},2T-1)$ for $N\ge 2T$.
 This number is called the
\emph{maximum-order complexity} (or \emph{nonlinear complexity}) of $\mathcal{S}$ and it is denoted by~$M(\mathcal{S})$. 
\aw{In other words, the maximum-order complexity is the length of a shortest (possibly nonlinear) feedback shift register.}
\aw{It is well-known that} $M(\mathcal{S})\le T-1$. 

In particular, restricting only to the homogeneous polynomials $f(X_1,\ldots,X_m)$ of degree one leads to the notion of the \emph{$N$th linear complexity} $L(\mathcal{S},N)$, 
the \emph{linear complexity} $L(\mathcal{S})=L(\mathcal{S},2T)\le T$ and  the \emph{linear complexity profile}  $(L(\mathcal{S},N))_{N\geq 1}$ of $\mathcal{S}$, respectively. See the list of surveys \cite{MW13,N2003,W2010}. 
The $N$th maximum-order complexity as well as the $N$th linear complexity are measures for the unpredictability of a sequence and thus its suitability in cryptography. \\

The $2$-adic complexity introduced by Goresky and Klapper \cite{GK2012,KG1997}
is closely related to the length of a shortest feedback with carry shift register (FCSR) which generates the sequence.
The theory of $2$-adic complexity has been very well developed for the periodic case.
More precisely, any $T$-periodic binary sequence $\mathcal{S}=(s_n)_{n\geq 0}$ uniquely corresponds to the rational number
\begin{equation}\label{rational}
\sum\limits_{n=0}^{\infty}s_n2^n=-\frac{\sum_{n=0}^{T-1}s_n2^n}{2^T-1}=-\frac{A}{q},
\end{equation}
where $0\leq A\leq q$, $\gcd(A,q)=1$ and
$$
q=\frac{2^T-1}{\gcd\left(2^T-1,\sum_{n=0}^{T-1}s_n2^n \right)},
$$
which is called the (minimal) \emph{connection integer} of $\mathcal{S}$ 
\cite{GK2012}.
Then the \emph{$2$-adic complexity} of $\mathcal{S}$, denoted by $\Phi_2(\mathcal{S})$,  is 
the binary logarithm $\log_2(q)$ of $q$.

In the aperiodic case, the \emph{$N$th $2$-adic complexity}, denoted by
$\Phi_2(\mathcal{S},N)$, is the binary logarithm of
$$
\min \left\{\max\{|f|,|q|\} : f,q\in\mathbb{Z}, q \mbox{ odd }, q\sum_{n=0}^{N-1}s_n2^n \equiv f \pmod{2^N}   \right\},
$$
see \cite[p.328]{GK2012} or \cite{W2023}.
\aw{It is trivial that $\Phi_2(\mathcal{S},N)\leq \Phi_2(\mathcal{S},N+1)$ for $N\geq 1$.}
The average behavior of the $2$-adic complexity and some asymptotic behavior of the $N$th $2$-adic complexity (more generally of the $d$-adic complexity of $d$-ary sequences, $d\geq 2$) are considered in Chapter 18.2 and Chapter 18.5 of \cite{GK2012}, respectively.

However, it seems that there are no results known on the relation between the $N$th $2$-adic complexity and other complexity measures.
Moreover, in contrast to the periodic case no results are known on the $N$th $2$-adic complexity of any attractive candidate for a pseudorandom sequence, \aw{that is a sequence with some (proved) desirable features and no known undesirable feature of pseudorandomness}.
Here we introduce the first theoretic method to study the aperiodic case, more precisely, we transfer \aw{some} known results on the $N$th maximum-order complexity to the $N$th $2$-adic complexity.
This leads to the main contribution of this work, that is, we will prove in Sections \ref{relation} and \ref{relper} the following inequalities
$$
 M(\mathcal{S},N)\le \left\lceil \Phi_2(\mathcal{S},N)\right\rceil+1, \quad N\geq 1,
$$
in the non-periodic case and
\begin{equation}\label{MOC-2AC}
M(\mathcal{S})\leq \lceil \Phi_2(\mathcal{S})\rceil
\end{equation}
if $\mathcal{S}$ is periodic, where $\left\lceil x  \right\rceil$ is the smallest integer $\ge x$.
The first inequality also implies a relation between the correlation measure of order $k$ and the $2$-adic complexity, see \aw{Corollary~\ref{corr}}.
Below we will also use $\left\lfloor x  \right\rfloor$ for the largest integer $\le x$.
We apply these bounds to several sequences including the Thue-Morse sequence along squares and the Legendre sequence.
 In addition, in the periodic case the idea of the proof of  Eq.(\ref{MOC-2AC}) leads to a characterization of the maximum-order complexity in terms of the rational number $-A/q$ defined by \eqref{rational}, which is stated in Subsection \ref{nece-suff}.
As a consequence, we prove  in Subsection \ref{ell-MOC} the maximality of the maximum-order complexity of binary $\ell$-sequences\aw{, which are those sequences for which the connection integer $q$ is a power of an odd prime such that $2$ is a primitive root modulo $q$, see \cite[Chapter 13]{GK2012}}. The result indicates that the bound in Eq.(\ref{MOC-2AC}) is sharp.
In Section \ref{max} we show that any periodic sequence with maximal maximum-order complexity has also maximal $2$-adic complexity.


We will use the notation
$f(N)=\mathcal O(g(N))$ if $|f(N)|\le cg(N)$ for some constant $c>0$ and
$f(N)=o(g(N))$ if $\lim\limits_{N\rightarrow \infty} \frac{f(N)}{g(N)}=0$.
Sometimes we also use $f(N)\ll g(N)$ and $g(N)\gg f(N)$ instead of $f(N)=\mathcal O(g(N))$.

\section{$N$th Maximum-order complexity and $N$th $2$-adic complexity}\label{relation}

In this section we prove a relation between the $N$th maximum-order complexity
$M(\mathcal{S},N)$ and the $N$th $2$-adic complexity $\Phi_2(\mathcal{S},N)$ 
and apply it to several prominent sequences.

\begin{theorem}\label{relation-N terms}
Let $\mathcal{S}=(s_n)_{n\geq 0}$ be a binary sequence. Then we have
$$
M(\mathcal{S},N)\le \left\lceil \Phi_2(\mathcal{S},N)\right\rceil +1
$$
for $N\geq 1$. 
\end{theorem}
Proof. Since otherwise the result is trivial we may assume
$$M(\mathcal{S},N)\geq 2$$ and 
put $m=M(\mathcal{S},N)-1$.
Then there exist $i,j$ with
\begin{equation}\label{jrange}
0\le i<j\le N-1-m
\end{equation}
and
\begin{equation}\label{distinct}
(s_i,s_{i+1},\ldots,s_{i+m-1})=(s_j,s_{j+1},\ldots,s_{j+m-1}),\quad s_{i+m}\not= s_{j+m},
\end{equation}
by \cite[Prop.~1]{jabo}, see also \cite[Thm.~2]{RK2005}.

Put
$$
S(2)=\sum_{n=0}^{N-1}s_n2^n
$$
and for $0\leq k<N$
$$
S_k(2)=\sum_{n=0}^{N-1-k}s_{n+k}2^n=2^{-k}\left(S(2)-\sum_{n=0}^{k-1}s_n2^n\right).
$$
Then we have for $i,j$ with $0\le i<j\le N-1-m$ chosen above to satisfy \eqref{distinct},
$$
S_i(2)\equiv \sum_{n=0}^m s_{n+i}2^n\equiv S_j(2)+2^m \pmod{2^{m+1}}
$$
and thus
\begin{eqnarray}\label{Sij(2)}
2^{j-i}\left(S(2)-\sum_{n=0}^{i-1}s_n2^n\right)
&\equiv& 2^jS_i(2)\equiv 2^jS_j(2)+2^{m+j}  \nonumber \\
&\equiv& S(2)-\sum_{n=0}^{j-1}s_n2^n+2^{m+j} \pmod{2^{m+j+1}}.
\end{eqnarray}

Note that $\Phi_2(\mathcal{S},N)\ge \Phi_2(\mathcal{S},m+j+1)$ by \eqref{jrange}
and assume 
$$
\Phi_2(\mathcal{S},m+j+1) =c
$$
for some $c\ge 0$, that is, there are 
integers $q$ and $h$, where $q$ is odd, with
\begin{equation}\label{hq}
\max\{|h|,|q|\}=  2^c
\end{equation}
and
\begin{eqnarray}\label{qS(2)=h}
qS(2)\equiv h\pmod{2^{m+j+1}}.
\end{eqnarray}
With \eqref{Sij(2)} multiplied by $q$ and \eqref{qS(2)=h}, we get
$$
2^{j-i}\left(h-q\sum_{n=0}^{i-1}s_n2^n\right)\equiv h-q\sum_{n=0}^{j-1}s_n2^n+2^{m+j}\pmod{2^{m+j+1}},
$$
that is
$$
2^{m+j}\equiv (2^{j-i}-1)h+q\left(\sum_{n=0}^{j-1}s_n 2^n-\sum_{n=0}^{i-1}s_n 2^{n+j-i}\right)\pmod{2^{m+j+1}}.
$$
\aw{Since
$$\left|\sum_{n=0}^{j-1}s_n 2^n-\sum_{n=0}^{i-1}s_n 2^{n+j-i}\right|\le \max\left\{\sum_{n=0}^{j-1}2^n,\sum_{n=0}^{i-1}2^{n+j-i}\right\}= 2^j-1$$
and by \eqref{hq}}
the right hand side is of absolute value at most
$$
2^{c+1}(2^j-1),
$$
which is not possible if $c\le m-1$. 
Hence, we get
$$
 \left\lceil \Phi_2(\mathcal{S},N)\right\rceil\aw{\ge \Phi_2(\mathcal{S},N)} \ge c > m-1=M(\mathcal{S},N)-2
$$
and the result follows.
\aw{Note that we cannot deduce $\Phi_2(\mathcal{S},N)\ge M(\mathcal{S},N)-1$ since $\Phi_2(\mathcal{S},N)$ may be no integer.}
\qed\\

According to Theorem \ref{relation-N terms}, a lower bound on  $M(\mathcal{S},N)$ implies a lower bound on~$\Phi_2(\mathcal{S},N)$.
Below we state several classes of sequences with known lower bounds on the $N$th maximum-order complexity which can now be interpreted as lower bounds on the $N$th $2$-adic complexity as well.

Pattern sequences (along squares/polynomial values):\\
For a positive integer $k$, the {\em pattern sequence} ${\cal P}_{k}=(p_n)_{n\geq 0}$ over ${\mathbb F}_2$ is defined by
$$
p_n=\left\{
\begin{array}{cl}
 p_{\lfloor n/2 \rfloor}+1, & \mbox{if } n\equiv -1 \pmod{2^k},\\
p_{\lfloor n/2 \rfloor},    & \mbox{otherwise},
\end{array}
\right.\quad n=1,2,\ldots
$$
with initial value $p_0=0$. Equivalently, $p_n$ is the parity of the number of occurrences of the all one pattern of length $k$ in the binary expansion of $n$.
For $k=1$ we get the \emph{Thue-Morse sequence} ${\cal T}=(t_n)_{n\geq 0}$
and for $k=2$ the {\em Rudin-Shapiro sequence} ${\cal R}=(r_n)_{n\geq 0}$.

We get $$\left\lceil \Phi_2(\mathcal{T},N)\right\rceil\ge \frac{N}{5},\quad N\ge 4,$$ 
\aw{and} 
$$\left\lceil \Phi_2(\mathcal{P}_k,N)\right\rceil \ge \frac{N}{6},\quad N\ge 2^{k+3}-7,\quad k\ge 2,$$  
\aw{from the lower bounds on the maximum-order complexity in} \cite{SW2019}.
Note that despite of a large $N$th maximum-order complexity and a large $N$th $2$-adic complexity, the pattern sequences ${\cal P}_k$ are highly predictable which can be measured in terms of a very small expansion complexity and a very large autocorrelation, see for example \cite{MW2022}. However, subsequences along polynomial values still keep the former desirable features but lose the latter undesirable ones.

For the {\em Thue-Morse sequence along squares} ${\cal T}'=(t_{n^2})_{n\geq 0}$ and  the {\em pattern sequence along squares} ${\cal P}'_{k}=(p_{n^2})_{n\geq 0}$,
we get 
$$\left\lceil \Phi_2(\mathcal{T}',N)\right\rceil \ge \sqrt{\frac{2N}{5}}-1,$$ 
and 
$$\left\lceil\Phi_2(\mathcal{P}'_k,N)\right\rceil\ge \sqrt{\frac{N}{8}}-1,\quad N\ge 2^{2k+2},\quad k\ge 2,$$ 
\aw{from the bounds in} \cite{SW19b}.

For $k\geq 1$ and the {\em pattern sequence along polynomial values of $f(X)$}, ${\cal P}''_{k}=(p_{f(n)})_{n\geq 0}$, where $f(X)\in \mathbb{Z}[X]$ is a monic polynomial
of degree $d\geq 2$ with $f(n)\geq 0$ for $n\geq 0$,
we get $\Phi_2(\mathcal{P}''_k,N)\gg N^{1/d}$, 
where the implied constants depend on $f(X)$ and $k$, see
\cite{P20}.
  
Sequence of the sum of digits in Zeckendorf base:  \\
 Let $F_0=0, F_1=1$ and $F_{i+2} = F_{i+1} +F_{i}$ for all $i\geq 0$, which forms the Fibonacci sequence. 
Each integer $n\geq 0$ can be represented uniquely by
$$
n = \sum_{i\geq 0}\varepsilon_i(n)F_{i+2},
$$
with $\varepsilon_i(n)\in\{0,1\}$ and $\varepsilon_i(n)\varepsilon_{i+1}(n)=0$ for all $i\geq 0$. Then the \emph{Zeckendorf base sum of digits function}
is defined by 
$$
s_Z(n)=\sum_{i\geq 0}\varepsilon_i(n), ~~n\geq 0.
$$
For the binary sequence of the Zeckendorf base sum of digits function ${\cal U}=(u_{n})_{n\geq 0}$ with $u_n=s_Z(n) \bmod 2$
and the binary sequence along polynomial values of the Zeckendorf base sum of digits function
${\cal U}'=(u_{f(n)})_{n\geq 0}$ with $u_{f(n)}=s_Z(f(n)) \bmod 2$,
where $f(x)\in \mathbb{Z}[x]$ is a monic polynomial
of degree $d\geq 2$ with $f(n)\geq 0$ for $n\geq 0$,
we get $\Phi_2(\mathcal{U},N)\gg N$ and $\Phi_2(\mathcal{U}',N) \gg N^{1/(2d)}$, see \cite{JPS21}.

For the Thue-Morse sequence, the Rudin-Shapiro sequence (both along the values of $f(X)\in\{X,X^2,X^3,X^4\}$) and the binary sequence of the Zeckendorf base sum of digits function\aw{,} we calculated the $N$th $2$-adic complexity up to $N=1\,000\,000$ which leads in all cases to the conjecture that $\Phi_2(\mathcal S,N)=\frac{N}{2}+\mathcal O(\log N)$.\\

We can also combine Theorem~\ref{relation-N terms} and \cite[Theorem~5]{CGGT2022} to get a lower bound on the $N$th $2$-adic complexity in terms of the $N$th correlation measure $C_2({\cal S},N)$ of order~$2$ introduced by Mauduit and S\'{a}rk\"{o}zy \cite{MS1997}. More precisely, for $k\geq 2$ the {\em $N$th correlation measure $C_k({\cal S},N)$ of order $k$} of $\mathcal{S}=(s_n)_{n\geq 0}$ is 
$$
C_k({\cal S},N)=\max_{U,D}\left|\sum_{i=0}^{U-1} (-1)^{s_{i+d_1}+ s_{i+d_2}+ \ldots +s_{i+d_k}}\right|,
$$
where the maximum is taken over all $D=(d_1,d_2,\ldots,d_k)$ and $U$
such that $0\le d_1<d_2<\cdots<d_k\le N-U$.
Then we have
$$C_2(\mathcal{S},N)\ge N-2^{M({\cal S},N)}+1$$
and the following result follows.

\begin{corollary}\label{corr}
Let $\mathcal{S}=(s_n)_{n\geq 0}$ be a binary sequence. Then we have
$$
\left\lceil\Phi_2({\cal S},N)\right\rceil \ge\log_{2}\left(N+1-C_2({\cal S},N)\right)-1.
$$
In particular, if $C_2({\cal S},N)=o(N)$, then we have
$$
\left\lceil\Phi_2({\cal S},N)\right\rceil\ge \log_2(N)-1+o(1).
$$
\end{corollary}

As an example we apply this relation to the Legendre sequence (along polynomial values):\\
For an odd prime $p$ and a squarefree polynomial $f(X)\in \mathbb{F}_p[X]$ of degree~$d$, 
the $p$-periodic Legendre sequence $\mathcal{L}=(\ell_n)_{n\geq 0}$ along the values of $f(X)$ is defined by
$$
\ell_n=\left\{
\begin{array}{cl}
1,  & \mbox{if }    \left( \frac{f(n)}{p}\right)=1,\\
0,  & \mbox{otherwise},
\end{array}
\right.
\quad n\geq 0, 
$$
where $\left( \frac{\cdot}{p}\right)$ is the Legendre symbol. 
By \cite{MS1997} we have 
$$
C_2(\mathcal{L},N)\ll dp^{1/2}\log p,\quad 1\le N\le p.
$$
and get 
$$
\Phi_2(\mathcal{L},N)\ge  \log_2(\min\{N,p\})-1+o(1)\quad \mbox{if } dp^{1/2}\log p=o(N).
$$

For the Legendre sequence with $f(X)=X$, it is conjectured that $\Phi_2(\mathcal L,N)=\min\{N/2,\Phi_2(\mathcal L)\}+\mathcal O(\log N)$, which we tested for all primes $p<50\,000$,
where $\Phi_2(\mathcal L)=\log_2(2^p-1)$, see \cite[Theorem~2]{HW2018}, \cite[Theorem~3]{XQL2014} and \cite{H2014}. 

\section{A relation between maximum-order complexity and $2$-adic complexity for periodic sequences}
\label{relper}

Now we turn to consider the case of periodic sequences. 
First we prove the following result, which is similar to the corresponding result for the maximum-order complexity of a $T$-periodic sequence $\mathcal{S}$, $M(\mathcal{S})=M(\mathcal{S},2T-1)$.

\begin{lemma}\label{2-adic N=2T}
Let $\mathcal{S}=(s_n)_{n\geq 0}$ be a binary sequence of period $T$. Then the $2$-adic complexity $\Phi_2(\mathcal{S})$  satisfies
 $$
 \Phi_2(\mathcal{S})=\Phi_2(\mathcal{S},2T+1)=\Phi_2(\mathcal{S},N)
 $$
 for any $N>2T$.
\end{lemma}
Proof. Let $\mathcal{S}$ correspond to the rational number
$$
\sum\limits_{n=0}^{\infty}s_n2^n=-\frac{\sum_{n=0}^{T-1}s_n2^n}{2^T-1}=-\frac{f}{q},
$$
with $0\leq f\leq q$ and $\gcd(f,q)=1$. It is clear that $q<2^{T}$ and $q$ is odd.

Assume $N> 2T$ and that there are integers $\widetilde{f}$ and odd $\widetilde{q}$ with
$$
\max\{|\widetilde{f}|,|\widetilde{q}|\}< q < 2^T
$$
and
$$
\widetilde{q} \sum_{n=0}^{N-1}s_n2^n\equiv \widetilde{f}  \pmod{2^{N}}.
$$
We obtain
$$
-\frac{f}{q}\equiv \frac{\widetilde{f}}{\widetilde{q}}  \pmod{2^{N}}, \mbox{ that is, } -\widetilde{q} f \equiv q \widetilde{f}  \pmod{2^{N}}.
$$
Since $|q\widetilde{f}+\widetilde{q}f|<2^{2T+1}\le 2^N$,
we derive $\widetilde{q} f + q \widetilde{f}=0$.
This leads to $q|\widetilde{q}$, which is impossible due to the assumption $|\widetilde{q}|< q$. 
Thus we get $\Phi_2(\mathcal{S},N)=\log_2(q)=\Phi_2(\mathcal{S})$ for $N> 2T$, which completes the proof. \qed

Note that we may have $\Phi_2(\mathcal{S},2T)<\Phi_2(\mathcal{S})$. For example, consider the $5$-periodic sequence starting with $(0,1,0,0,1)$ which is the binary expansion of $18$.
Since $\gcd(2^5-1,18)=1$ we get $\Phi_2(\mathcal{S})=\log_2(31)$. However, we have
$$19\cdot 18\cdot (1+2^5)\equiv 22 \pmod{2^{10}},$$
may take $q=19$ and $f=22$
and thus get
$$\Phi_2(\mathcal{S},10)\le \log_2(22)<\Phi_2(\mathcal{S}).$$

So Theorem \ref{relation-N terms} and Lemma \ref{2-adic N=2T} indicate
\begin{eqnarray*}
M(\mathcal{S})&=&
M(\mathcal{S},2T-1)\\ 
&\le& 
\left\lceil \Phi_2(\mathcal{S},2T-1)\right\rceil +1\le
\left\lceil \Phi_2(\mathcal{S},2T+1)\right\rceil +1 = \left\lceil \Phi_2(\mathcal{S})\right\rceil +1.
\end{eqnarray*}
Below we prove a slightly stricter bound in another way (for the periodic case).

\begin{theorem}\label{new-upperB}
Let $\mathcal{S}=(s_n)_{n\geq 0}$ be a binary sequence of period $T$.
Then we have
$$
M(\mathcal{S})\leq \left\lceil \Phi_2(\mathcal{S})  \right\rceil.
$$
\end{theorem}
Proof. According to \cite[Prop.~2]{R2006}, we will compute the
minimum integer $k$, which equals to the maximum-order complexity $M(\mathcal{S})$,
such that all ($T$ many)
subsequences of length $k$:
$$
(s_{0},s_{1},\ldots,s_{k-1}), ~~(s_{1},s_{2},\ldots,s_{k}),\ldots,  (s_{T-1},s_{0},\ldots,s_{k-2})
$$
are distinct.

Let
$$
\sum\limits_{n=0}^{\infty}s_n2^n=-\frac{\sum_{n=0}^{T-1}s_n2^n}{2^T-1}=-\frac{f}{q},
$$
with $0\leq f\leq q$ and $\gcd(f,q)=1$. Then $\Phi_2(\mathcal{S})=\log_2(q)$. Below we prove the statement in terms of $\log_2(q)$.

If $T=1$, that is, $\mathcal{S}$ is constant, 
we derive $q=1$ and $M(\mathcal{S})=0 = \log_2(1) = \Phi_2(\mathcal{S})$.

Below we suppose $T\geq 2$ and in this case we have $0<f<q$. Now we assume that $T>\left\lceil \log_2(q) \right\rceil$,
since otherwise the result is trivial by
$$\lceil\log_2(T)\rceil\leq M(\mathcal{S})<T,$$
see \cite[Prop.~2]{jabo}.
For $0\leq \tau<T$, suppose that the cyclic (left) $\tau$-shift of $\mathcal{S}$, denoted by $\mathcal{S}^{(\tau)}$,
corresponds to the rational number $-\frac{h_{\tau}}{q}$ with $0<h_{\tau}<q$ and $\gcd(h_{\tau},q)=1$. It is clear that $h_0=f$.
Among these $T$ many shift sequences, we count the ones with the same beginning $N$ terms.

If $\mathcal{S}^{(i)}$ (associated to $-\frac{h_i}{q}$) and $\mathcal{S}^{(j)}$  (associated to $-\frac{h_j}{q}$) are with the same beginning $N$ terms, we have
$$
-\frac{h_i}{q}\equiv  -\frac{h_j}{q} \pmod{2^N},
$$
which holds if and only if $h_i\equiv h_j \pmod{2^N}$, since $q|(2^T-1)$.

Let $2^{m-1}<q<2^m$ for some positive integer $m$, so $m=\left\lceil \log_2(q) \right\rceil$.  If we choose $N=m$,
we will find that
$h_i\equiv h_j \pmod{2^m}$  if and only if $h_i=h_j$ since $0<h_i, h_j<q<2^m$. It means that the beginning $m$ terms of $-\frac{h_i}{q}$ are different from the ones of
 $-\frac{h_j}{q}$ for all $0\leq i<j<T$.
Then we derive that any subsequences $(s_i,s_{i+1},\ldots,s_{i+m-1})$ of length $m$ for $0\leq i<T$ are distinct, and hence
$M(\mathcal{S})\leq m=\left\lceil \log_2(q) \right\rceil$.
Finally, together with the notion of the $2$-adic complexity, we get $M(\mathcal{S})\leq\left\lceil \Phi_2(\mathcal{S})  \right\rceil$ directly.  \qed

As far as we know, these are the first results on the relation between the maximum-order complexity and the $2$-adic complexity.
It disproves a claim by Goresky and Klapper ``If a sequence is generated by an FCSR with nonnegative memory, then its $N$-adic span is no greater than its  maximum order complexity" in \cite[p.329]{GK2012}. For the requirement of nonnegative memory, see \cite[Prop.4.7.1]{GK2012}.

We remark that it is also important 
to consider the \emph{symmetric $2$-adic complexity} of~$\mathcal{S}$, which is the minimum of the $2$-adic complexities of $\mathcal{S}$ and $\mathcal{S}^{rev}$, where $\mathcal{S}^{rev}$ is the sequence formed by reversing each period of $\mathcal{S}$, see  \cite[Sect. 16.2]{GK2012}, since $\Phi_2(\mathcal{S}^{rev})$ may be substantially smaller than $\Phi_2(\mathcal{S})$. By Theorem \ref{new-upperB} we have $M(\mathcal{S}^{rev})\leq \left\lceil \Phi_2(\mathcal{S}^{rev}) \right\rceil$. By \cite[Prop.2]{R2006} it is clear that $M(\mathcal{S}^{rev})=M(\mathcal{S})$ and so 
$$M(\mathcal{S})\leq \min(\left\lceil \Phi_2(\mathcal{S})\right\rceil, \left\lceil \Phi_2(\mathcal{S}^{rev}) \right\rceil).$$

\section{Maximum-order complexity of periodic sequences}

\subsection{A characterization of the maximum-order complexity}\label{nece-suff}

The idea in the proof of Theorem \ref{new-upperB} helps us to characterize the maximum-order complexity of periodic sequences.

\begin{theorem}\label{nece-suff cond}
Let $\mathcal{S}=(s_n)_{n\geq 0}$ be a binary sequence of period $T (\geq 2)$.
If 
$$\sum_{n=0}^\infty s_n 2^n=-A/q \quad \mbox{with }0< A< q \mbox{ and }\gcd(A,q)=1$$ and 
$$
D_A=\{0\leq u<q : u\equiv A\cdot 2^n \pmod{q}, 0\leq n<T\},
$$
then $M(\mathcal{S})=N$ if and only if $N$ is the least integer such that
$$
u\not\equiv v \pmod{2^N}
$$
for any different $u, v\in D_A$.
\end{theorem}
Proof. \aw{Suppose that  $\mathcal{S}^{(\tau)}$, the (left) $\tau$-shift of $\mathcal{S}$ for $0\leq \tau<T$, corresponds to the rational number $-\frac{A^{(\tau)}}{q}$. We see that 
 $A^{(\tau)}\equiv A2^{T-\tau} \pmod{q}$ and hence 
  $D_A=\{A^{(\tau)}:0\le \tau<T\}$.}

For $N\geq 1$, the first $N$ elements of $\mathcal{S}^{(i)}$ are the same as the ones of $\mathcal{S}^{(j)}$ for $0\leq i<j<T$ if and only if
$$
-\frac{A^{(i)}}{q} \equiv - \frac{A^{(j)}}{q} \pmod{2^N},
$$
which holds if and only if $A^{(i)}\equiv A^{(j)} \pmod{2^N}$, since $q$ is odd.
\aw{This means that for $u, v\in D_A$, 
if $u\not\equiv v \pmod{2^N}$, we derive that 
the first $N$ elements of $-u/q$ are different from the ones of $-v/q$, which completes} 
the proof. \qed \\

The set $\langle 2\rangle=\{0\leq u<q : u= 2^n \pmod{q}, 0\leq n<T\}$ generated by $2$ modulo $q$ is a sub-group of $\mathbb{Z}_q^*=\{0<u<q : \gcd(u,q)=1\}$
under integer multiplication modulo $q$. Thus according to Theorem \ref{nece-suff cond},  to analyze the maximum-order complexity of periodic sequences, one only needs to consider the partition
$$
\mathbb{Z}_q^{*}= g_1\langle 2\rangle \cup g_2\langle 2\rangle \cup\cdots \cup g_{\varphi(q)/T}\langle 2\rangle,
$$
a union of co-sets of $\langle 2\rangle$,
where $g_i\langle 2\rangle=\{g_iu \pmod{q}: u\in\langle 2\rangle\}\subseteq \mathbb{Z}_q^{*}$ for $1\leq i\leq \varphi(q)/T$. We note that
$D_A$ in Theorem \ref{nece-suff cond} is a co-set of $\langle 2\rangle$.

\subsection{Maximum-order complexity of $\ell$-sequences}\label{ell-MOC}

As a consequence, we consider the case when $D_A=\mathbb{Z}_q^{*}$. We find that $2$ modulo $q$ is primitive and hence $\mathcal{S}$ is an $\ell$-sequence in this case.
In particular, $q$ has to be a power of an odd prime.
The following theorem will indicate that the bound in Theorem \ref{new-upperB} is sharp.

\begin{lemma}\label{lem-359}
Suppose that $q=p^r$ the power of an odd prime for $r\geq 1$ and $2$ modulo $q$ is primitive. If $q$ is of the form $2^k+1$ for some integer $k\geq 1$,
 then the possible value $q$ is in $\{3,5,9\}$.
\end{lemma}
Proof. It is clear that
$$
q=\left\{
\begin{array}{cl}
3,  & \mbox{ if } k=1, \\
5,  & \mbox{ if } k=2, \\
3^2,  & \mbox{ if } k=3,
\end{array}
\right.
$$
which satisfies all other assumptions in the lemma.
Now we consider the case $k\geq 4$ (and hence $q\geq 17$).

If $r=1$, that is, $q$ is an odd prime, we get
$$
\left( \frac{2}{q} \right)=-1= (-1)^{(q^2-1)/8},
$$
since  $2$ modulo $q$ is primitive, where $\left( \frac{\cdot}{\cdot} \right)$ is the Legendre symbol. We derive
$q\equiv \pm 3 \pmod{8}$, which contradicts $q=2^k+1 \equiv 1 \pmod{8}$.

If $r\geq 2$, we see that
$$
2^k\equiv -1 \pmod{q} ~~\mbox{ and } 2^{\varphi(q)/2}\equiv -1 \pmod{q},
$$
the latter holds due to $2$ modulo $q$ being primitive. Hence $k=c \varphi(q)/2$ for some odd positive integer $c\geq 1$.

We see that $2$ modulo $p^\ell$ is primitive for all $1\leq \ell\leq r$ due to $2$ modulo $q$ being primitive again.
So we have $2^{\varphi(p^{r-1})/2}\equiv -1 \pmod{p^{r-1}}$ and write
 $$
2^{p^{r-2}(p-1)/2}= wp^{r-1}-1
$$
for some positive integer $w\geq 1$. We compute
$$
q=2^k+1=2^{c\varphi(q)/2}+1=(2^{p^{r-2}(p-1)/2})^{cp}+1=(wp^{r-1}-1)^{cp}+1.
$$
Together with $wp^{r-1}-1 =wp^{1/3}p^{r-4/3}-1> p^{r-4/3}$ and $cp(r-4/3)\geq 3(r-4/3)=r+(2r-4)\geq r$ (since $q\geq 17$ and $r\geq 2$), we derive
$$
(wp^{r-1}-1)^{cp}+1 > (p^{r-4/3})^{cp} +1 >p^r=q,
$$
a contradiction.

Putting everything together, we see that only $q\in\{3,5,9\}$ satisfies the requirements. \qed

\begin{theorem}
Let $\mathcal{S}=(s_n)_{n\geq 0}$ be a binary $\ell$-sequence with (minimal) connection integer $q(\geq 3)$, which is an odd prime power.
Then the maximum-order complexity $M(\mathcal{S})$ of $\mathcal{S}$ satisfies
$$
M(\mathcal{S})=\left\{
\begin{array}{ll}
\left\lfloor \log_2(q) \right\rfloor, & \mbox{if } q\in \{3,5,9\},\\
\left\lceil \log_2(q) \right\rceil,   & \mbox{otherwise}.
\end{array}
\right.
$$
\end{theorem}
Proof. We assume that $\mathcal{S}$ is an $\ell$-sequence defined by Eq.(\ref{FCSR-seq}) with some integer $A$. Since 2 modulo $q$ is primitive,
we see that $D_A=\mathbb{Z}_q^*$ defined in Theorem \ref{nece-suff cond}.

\begin{itemize}
  \item If $q=3$, we see that the $\ell$-sequence $\mathcal{S}=(10)$ or $(01)$, whose period is $2$.
We check that $M(\mathcal{S})=1=\left\lfloor \log_2(q) \right\rfloor$ by \cite[Prop.~2]{R2006}.

Below we consider the case of $q\geq 5$.
Let $2^{m-1}<q<2^m$ for some positive integer $m\geq 3$.

  \item  If $q=2^{m-1}+1$, we find that $x\leq q-1=2^{m-1}$ for any $x\in D_A$ and hence
all $x\in D_A$ modulo $2^{m-1}$ are distinct. However, we have $1+2^{m-2}\neq 1$ since $m\geq 3$ and
$$
1+2^{m-2}\equiv 1 \pmod{2^{m-2}}.
$$
We remark that both $1$ and $1+2^{m-2}$ are in $\mathbb{Z}_q^{*}(=D_A)$. So by  Theorem \ref{nece-suff cond} we derive
that $M(\mathcal{S})=m-1=\left\lfloor \log_2(q) \right\rfloor$. Furthermore, we have $q=5$ if $m=3$ and $q=9$ if $m=4$.
But if $m\geq 5$, no such $q$ exists by Lemma \ref{lem-359}.

  \item If $q>2^{m-1}+1$, we find that $q-1> 2^{m-1}$ and all $x\in D_A$ modulo $2^{m}$ are distinct.
However, we have $1+2^{m-1}\neq 1$ and $2+2^{m-1}\neq 2$  since $m\geq 3$ and
$$
1+2^{m-1}\equiv 1 \pmod{2^{m-1}} \mbox{ and } 2+2^{m-1}\equiv 2 \pmod{2^{m-1}}.
$$
We remark that either both $1$ and $1+2^{m-1}$ or both $2$ and $2+2^{m-1}$ are in $\mathbb{Z}_q^{*}(=D_A)$
 \footnote{We need to prove either $\gcd(1+2^{m-1},q)=1$ or $\gcd(2+2^{m-1},q)=1$. Let $q=p^r$, an odd prime-power. If $p|(1+2^{m-1})$, then $p\nmid(2+2^{m-1})$.}.
 So by  Theorem \ref{nece-suff cond} we derive
that $M(\mathcal{S})=m=\left\lceil \log_2(q) \right\rceil$.  \qed

\end{itemize}

We list some examples of $\ell$-sequences in Table \ref{Tab-ell}.

\begin{table}[h]
\centering
\begin{tabular}{ccccc}
\hline\noalign{\smallskip}
$2^{m-1}<q<2^m$         &  $T=\varphi(q)$  & $\left\lceil \log_2(q) \right\rceil$   & $M(\mathcal{S})$     & Remarks    \\
 \noalign{\smallskip} \hline \noalign{\smallskip}
$2<q=3<2^2$             &   $2$            &   $2$                                  & $1$                  &  $M(\mathcal{S})=\left\lfloor \log_2(q) \right\rfloor$    \\
$2^3<q=3^2<2^4$         &   $6$            &   $4$                                  & $3$                  &  $M(\mathcal{S})=\left\lfloor \log_2(q) \right\rfloor$    \\
$2^4<q=3^3<2^5$         &   $18$           &   $5$                                  & $5$                  &         \\
$2^2<q=5<2^3$           &   $4$            &   $3$                                  & $2$                  &  $M(\mathcal{S})=\left\lfloor \log_2(q) \right\rfloor$     \\
$2^9<q=5^4<2^{10}$      &   $500$          &   $10$                                 & $10$                 &         \\
$2^4<q=19<2^5$          &   $18$           &   $5$                                  & $5$                  &          \\
$2^8<q=19^2<2^9$        &   $342$          &   $9$                                  & $9$                  &         \\
$2^{12}<q=19^3<2^{13}$  &   $6498$         &   $13$                                 & $13$                 &        \\
\hline
\end{tabular}
\caption{$M(\mathcal{S})$ of binary $\ell$-sequence $\mathcal{S}$ with connection integer $q$}\label{Tab-ell}
\end{table}

For non-$\ell$-sequences, we have a different phenomenon, see examples in Table \ref{Tab-non-ell}.

\begin{table}[h]
\centering
\begin{tabular}{cccc}
\hline\noalign{\smallskip}
$2^{m-1}<q<2^m$         &  $T=\mathrm{ord}_q(2)\neq \varphi(q)$       & $\left\lceil \log_2(q) \right\rceil$   & $M(\mathcal{S})\in$      \\
 \noalign{\smallskip} \hline \noalign{\smallskip}
$2^5<q=51<2^6$          &   $8$                              &   $6$                                  & $\{4,5\}$               \\
$2^5<q=63<2^6$          &   $6$                              &   $6$                                  & $\{3,4,5\}$             \\
$2^6<q=65<2^7$          &   $12$                             &   $7$                                  & $\{4,6\}$                \\
$2^6<q=93<2^7$          &   $10$                             &   $7$                                  & $\{4,5,6\}$             \\
$2^7<q=217<2^8$         &   $15$                             &   $8$                                  & $\{5,6,7,8\}$            \\
\end{tabular}
\caption{$M(\mathcal{S})$ of $\mathcal{S}$ with $s_n= (A2^{-n} \bmod{q}) \bmod 2$ for different $A$}\label{Tab-non-ell}
\end{table}

\subsection{Binary sequences $\mathcal{S}$ of period $T$ with 
$M(\mathcal{S})=T-1$
}\label{max}

Binary sequences $\mathcal{S}$ of period $T$ with $M(\mathcal{S})=T-1$ were considered before in \cite{R2005,R2006,PM2008,SZLH2017}.
Now we look at them in another way.
Suppose that  $\mathcal{S}$ corresponds to the  rational number $\frac{A}{q}$ with $-q<A<0$ and $\gcd(A,q)=1$. Then by \cite[Thm. 4.5.2]{GK2012}, $\mathcal{S}$ can be defined by 
\begin{equation}\label{FCSR-seq}
s_n=(A2^{-n} \bmod q) \bmod 2, ~~~ n\geq 0.
\end{equation}
Below we list some examples of $\mathcal{S}$ with $M(\mathcal{S})=T-1$.
\begin{itemize}
  \item  If we choose $q=31(=2^5-1)$ in Eq.(\ref{FCSR-seq}), we produce $\mathcal{S}=(11000)$ with period $T=5$ if $A=3$ and  check $M(\mathcal{S})=3(<T-1)$.
While if $A=5$ we produce $\mathcal{S}=(10100)$ and check $M(\mathcal{S})=4(=T-1)$.

  \item  If we choose $q=127(=2^7-1)$ and $A=37$ in Eq.(\ref{FCSR-seq}), we produce $\mathcal{S}=(1010010)$ with period $T=7$.
We check that $M(\mathcal{S})=6(=T-1)$.

  \item  If we choose $q=255(=2^8-1)$ and $A=173$ in Eq.(\ref{FCSR-seq}), we produce $\mathcal{S}=(10110101)$ with period $T=8$.
We check that $M(\mathcal{S})=7(=T-1)$.
\end{itemize}

For such $\mathcal{S}$, 
\aw{by Theorem \ref{new-upperB} we see that $\lceil\Phi_2(\mathcal{S})\rceil\geq T-1$ and so $$\Phi_2(\mathcal{S})\in\left\{\log_2\left(\frac{2^T-1}{3}\right),~\log_2\left(2^T-1\right) \right\}.$$}
\aw{However, }we prove that their $2$-adic complexity is maximal.

\begin{theorem}
Let $\mathcal{S}=(s_n)_{n\geq 0}$ be a binary sequence of period $T\geq 2$. If $M(\mathcal{S})=T-1$, then the $2$-adic complexity of $\mathcal S$ is also maximal, that is, 
$\Phi_2(\mathcal{S})=\log_2(2^T-1)$.
\end{theorem}
Proof.
Since the maximum-order complexity is the same for every shift of $\mathcal S$, we may assume that there exists an integer $d\in \{1,\ldots,T-1\}$ such that
\begin{align*}
s_i&=s_{i+d}\quad\text{for }0\le i\le T-3,\\
	s_i&=1-s_{i+d}\quad \text{for }i\in\{T-2,T-1\},
\end{align*}
see Equation (2) in \cite[Sect.III, A, p.6190]{SZLH2017}. We have $\gcd(d,T)=1$ by \cite[Prop.~1]{SZLH2017} and hence $e=d^{-1}\pmod T$ exists. 

Using the two equations above, we can check that 
$$
s_{id-1}=1-s_{T-1} \mbox{ for }   1\le i \le T-e, ~~~
s_{jd-2}=1-s_{T-2}  \mbox{ for } 1\le j \le e.
$$
We also have
$$
s_{(T-e)d-1}=s_{T-2},\quad 
 s_{ed-2}=s_{T-1}\quad \mbox{and}\quad s_{T-1}=1-s_{T-2}.
$$
We note that 
$$
\{id-1 \pmod{T}: 1\le i\le T-e\} \cup \{jd-2 \pmod{T}: 1\le j\le e\}
=\{0,1,\ldots,T-1\}.
$$

In the case $(s_{T-2},s_{T-1})=(0,1)$ we get 
\[
S(2)\equiv\sum_{j=1}^e 2^{jd-2}\equiv 2^{d-2} (2^d-1)^{-1}(2^{ed}-1)\equiv 2^{d-2}(2^d-1)^{-1}\pmod {2^T-1},
\]
\aw{where we used
$$ed\equiv 1\pmod T \quad \mbox{and thus} \quad 2^{ed}\equiv 2\pmod {2^T-1}$$
in the last step,}
from which we derive 
\[
\gcd(S(2),2^T-1)=1.
\]
The case $(s_{T-2},s_{T-1})=(1,0)$ can be treated in a similar way. 

Hence the connection integer of $\mathcal{S}$ is $2^T-1$ and the $2$-adic complexity of $\mathcal S$ is maximal, which completes the proof. \qed

\aw{For sequences $\mathcal S$ with $M(\mathcal S)=T-2$, which are characterized in \cite{XZLJ2019}, the
2-adic complexity is however not necessarily maximal anymore. For
example for the sequence $\mathcal S$ starting with $(0, 0, 1, 0, 0,
1, 0, 0) $ of period $T=8$, we have $M(\mathcal S)=T-2$ but
$\Phi_2(\mathcal S)=\log_2(\frac {2^T-1}3)$. }

\section{Final remarks and conclusions}


We have discussed the relationship between the maximum-order complexity and the $2$-adic complexity for any binary sequence. \aw{More precisely, the $2$-adic complexity is at least of the order of magnitude of the maximum-order complexity. If the order of magnitude of the maximum-order complexity is maximal, then our bound is essentially tight. However, for a typical sequence the maximum-order complexity is much smaller than the $2$-adic complexity.}

There is another complexity measure called the expansion complexity introduced by
Diem \cite{di12}. Let $G(x)=\sum_{i\geq 0}s_i x^i$
be the {\em generating function} of $\mathcal{S}=(s_n)_{n\geq 0}$, which is viewed as a formal power series over $\mathbb{F}_2$.
The {\em $N$th expansion complexity}
$E({\cal S},N)$ is $0$ if $s_0=\ldots=s_{N-1}=0$ and otherwise the least total degree
of a nonzero polynomial $h(x,y)\in \mathbb{F}_2[x,y]$ with
$$
h(x,G(x))\equiv 0 \bmod x^N.
$$
Note that $E({\cal S},N)$ depends only on the first $N$ terms of ${\cal S}$ and it has an expected value of order of magnitude $N^{1/2}$, see \cite[Theorem 2]{GM2020}.
The sequence $(E(\mathcal{S},N))_{N\geq 1}$
is referred to as the \emph{expansion complexity profile} of $\mathcal{S}$.
The value 
$$
E({\cal S})=\sup_{N\ge 1} E({\cal S},N)
$$
is the \emph{expansion complexity} of $\mathcal{S}$, see \cite{GM2020,GMH2018,HW2017} for more details on the expansion complexity of sequences.

\aw{Another measure of pseudorandomness, the \emph{rational complexity} $R({\cal S},N)$ (resp.\ $R(\cal S)$ in the periodic case), was introduced and studied in \cite{VCC2022} dealing with so-called F$\mathbb{Q}$SRs instead of LFSRs (linear complexity) or FCSRs ($2$-adic complexity). }

Finally, we \aw{give a list of the relationships between five} complexities. For periodic sequences $\mathcal{S}$, we have

\begin{itemize}
  \item $M(\mathcal{S}) \leq L(\mathcal{S}) = E(\mathcal{S})-1$.
  
  \item $M(\mathcal{S}) \le \left\lceil\Phi_2(\mathcal{S})\right\rceil$ (our result).
  

  \item The linear complexity and the $2$-adic complexity complement each other. 
For example, for a binary $m$-sequence $\mathcal{M}=(m_n)_{n\ge 0}$ of period $T=2^{r}-1$, that is,
$$m_n={\rm Tr}(g^n)=\sum_{j=0}^{r-1}g^{2^jn},\quad n\ge 0,$$
where $g$ is a primitive element of $\mathbb{F}_{2^r}$ and Tr is the absolute trace of $\mathbb{F}_{2^r}$,
we see that
$L(\mathcal{M})=r$, which is minimal for any sequence of least period $2^r-1$, but $\Phi_2(\mathcal{M})=\log_2(2^T-1)$, which is the maximum, see \cite{TQ2010}.
For a binary $\ell$-sequence $\mathcal{S}$ with minimal connection integer $q$ of period $T=\varphi(q)$, where $\varphi$ is Euler's totient function, we see that
$L(\mathcal{S})\leq (q+1)/2$, the bound of which is sharp, for example if $p$ and $q=2p+1$ are primes with 2 being primitive modulo $p$ and modulo $q$ respectively,
then the linear complexity is $L(\mathcal{S})=(q+1)/2~(=p+1)$ \cite{SL2000}. But $\Phi_2(\mathcal{S})=\log_2(q)$, which is small.
\item \aw{$R(\mathcal{S})=\Phi_2(\mathcal{S}^{rev})$ by \cite[Theorem 13]{VCC2022}.} 
\item \aw{$M(\mathcal{S})  \le \left\lceil \Phi_2(\mathcal{S}^{rev})\right\rceil$.} 
\end{itemize}

In the aperiodic case, we have the \aw{following} relationships between the \aw{$N$th complexities for} $\mathcal{S}$:

\begin{itemize}
  \item $M(\mathcal{S},N) \leq L(\mathcal{S},N)$. Hence
\aw{a large $N$th maximum-order complexity implies a large $N$th linear complexity.}
However, $M(\mathcal{S},N)$  should not be
too large since otherwise the correlation measure of order $2$ is large, see \cite{MW2022}.
In particular, the expected value of $M(\mathcal{S},N)$ is of order of magnitude $\log(N)$ \cite{jabo} and the expected value of $L(\mathcal{S},N)$ is $\frac{N}{2}+\mathcal O(1)$ \cite{N1988}.
    
  \item $E(\mathcal{S},N) \leq L(\mathcal{S},N)+1$. In fact it was proved in \cite{MNW2017} that
$$E({\cal S},N)\leq \min\{L({\cal S},N)+1, N+2-L({\cal S},N)\}.$$
   
  \item $M(\mathcal{S},N) \le \left\lceil\Phi_2(\mathcal{S},N)\right\rceil+1$  (our result).
  
  \item  The $N$th expansion complexity and the $N$th maximum-order complexity complement each other. We note that the expected value of $M(\mathcal{S},N)$ is of order of magnitude $\log(N)$ and 
   $E({\cal S},N)$ has an expected value of order of magnitude $N^{1/2}$. However, the pattern sequences $\mathcal{P}_k$ of length $k$ including the Thue-Morse sequence 
  $\mathcal{T}$ (see notions in Section \ref{relation}) have bounded expansion complexity, that is $E(\mathcal{P}_k,N)\leq 2^k+3$ for $k\geq 1$, see \cite{MW2022}. While the maximum-order complexity is of the largest possible order of magnitude $N$, that is $M(\mathcal{T},N)> N/5$ for $N>5$ and   $M(\mathcal{P}_k,N)> N/6$ for $N\geq 2^{k+3}-7$ if $k\geq 2$, see \cite{SW2019}. 

  \item \aw{\cite{VCC2022} provides examples that rational complexity, $2$-adic complexity and linear complexity complement each other. More precisely, in these examples each two of these measures are close to the expected value whereas the third one is much smaller and only one of these three measures detects the non-randomness.}
\end{itemize}


\begin{thebibliography}{00}


 \bibitem{CGGT2022}
 Z. Chen, A. G\'{o}mez, D. G\'{o}mez-P\'{e}rez, A. Tirkel.
Correlation measure, linear complexity and maximum order complexity for families of binary sequences. Finite Fields Appl. 78: 101977 (2022)

 \bibitem{di12}
  C. Diem. On the use of expansion series for stream ciphers. LMS J. Comput. Math. 15: 326-340 (2012)






\bibitem{GM2020}
D. G\'{o}mez-P\'{e}rez, L. M\'{e}rai. Algebraic dependence in generating functions and expansion complexity. Adv. Math. Commun. 14(2): 307-318 (2020)

\bibitem{GMH2018}
D. G\'{o}mez-P\'{e}rez, L. M\'{e}rai, H. Niederreiter. On the expansion complexity of sequences over finite fields. IEEE Trans. Inform. Theory 64(6): 4228-4232 (2018)




\bibitem{GK2012}
M. Goresky, A. Klapper. Algebraic Shift Register Sequences. Cambridge University Press, Cambridge (2012)



\bibitem{HW2017}
R. Hofer and A. Winterhof.
 Linear complexity and expansion complexity of some number theoretic sequences.
 In: Arithmetics in Finite Fields-WAIFI2016. Lect. Notes Comput. Sci.  10064, 67-74, Springer-Verlag, Cham (2017)


\bibitem{HW2018} R. Hofer and A. Winterhof. On the $2$-adic complexity of the two-prime generator.
 IEEE Trans. Inform. Theory  64(8): 5957--5960 (2018)

\bibitem{H2014} H. Hu. Comments on `A new method to compute the $2$-adic complexity of binary sequences'. IEEE Trans. Inform. Theory 60(9): 5803--5804 (2014) 


\bibitem{JPS21}
D. Jamet, P. Popoli, T. Stoll. Maximum order complexity of the sum of digits function in Zeckendorf base and polynomial subsequences.
 Cryptogr. Commun.  13(5):  791-814  (2021)
		

 \bibitem{ja89}
 C. J. A. Jansen. Investigations on nonlinear streamcipher systems: construction and evaluation methods. Ph.D. dissertation, Technical University of Delft, Delft (1989)

\bibitem{jabo}
C. J. A. Jansen, D. E. Boekee. The shortest feedback shift
register that can generate a given sequence. In G. Brassard (Ed.): Advances in
Cryptology--CRYPTO'89, Lect. Notes Comput. Sci. 435: 90-99,
Springer-Verlag, Berlin Heidelberg  (1990)


\bibitem{ja91} C. J. A. Jansen. The maximum order complexity of sequence ensembles.
In D.W. Davies (Ed.): Advances in Cryptology - EUROCRYPT'91, Lect. Notes Comput. Sci. 547: 153-159,
Springer-Verlag, Berlin Heidelberg (1991)



\bibitem{KG1997}
 A. Klapper, M. Goresky.
 Feedback shift registers, $2$-adic span and combiners with memory. J. Cryptol. 10(2): 111-147 (1997)



 \bibitem{LKK2007}
 K. Limniotis, N. Kolokotronis, N. Kalouptsidis.
 On the nonlinear complexity and Lempel-Ziv complexity of finite length
 sequences. IEEE Trans. Inform. Theory 53(11): 4293-4302 (2007)

\bibitem{MS1997}
C. Mauduit, A. S\'{a}rk\"{o}zy. On finite pseudorandom binary sequences I: measure of pseudorandomness,
the Legendre symbol. Acta Arith. 82(4): 365-377  (1997)
		


 \bibitem{MW13}
W. Meidl, A. Winterhof. Linear complexity of sequences and multisequences.
  In: G. L. Mullen, D. Panario (eds.): Handbook of Finite Fields, 324-336, CRC Press, Boca Raton, FL (2013)


\bibitem{MW2022}
L. M\'{e}rai, A. Winterhof. Pseudorandom sequences derived from automatic sequences. Cryptogr. Commun. 14(4): 783-815 (2022)


\bibitem{MNW2017}
L. M\'{e}rai, H. Niederreiter, A. Winterhof. Expansion complexity and linear complexity of sequences
over finite fields. Cryptogr. Commun. 9(4): 501-509 (2017)

\bibitem{N1988}
 H. Niederreiter. The probabilistic theory of linear complexity. In: Advances in Cryptology — EUROCRYPT'88, Lecture
Notes in Comput. Sci. 330: 191-209, Springer-Verlag, Berlin Heidelberg  (1988)

\bibitem{N2003}
H. Niederreiter. Linear complexity and related complexity measures for sequences.
In: T. Johansson, S. Maitra (eds.): Progress in Cryptology - INDOCRYPT 2003,  Lect. Notes Comput. Sci. 2904: 1-17, Springer-Verlag, Berlin Heidelberg  (2003)

 \bibitem{nixi14}
	H. Niederreiter, C. Xing.
Sequences with high nonlinear complexity. IEEE Trans. Inform. Theory 60(10): 6696-6701 (2014)

\aw{
\bibitem{PM2006}
G. Petrides, J. Mykkeltveit.
On the classification of periodic binary sequences into nonlinear complexity classes. 
In: G. Gong, T. Helleseth, H. Song, K. Yang(eds) Sequences and Their Applications – SETA 2006, Lect. Notes Comput. Sci. 4086: 209-222, Springer-Verlag, Berlin Heidelberg (2006)  
}

 \bibitem{PM2008}
G. Petrides, J. Mykkeltveit.
Composition of recursions and nonlinear complexity of periodic binary sequences. Des. Codes Cryptogr. 49(1-3): 251-264 (2008)



\bibitem{P20} P. Popoli. On the maximum order complexity of Thue-Morse and Rudin-Shapiro sequences along polynomial values.
 Unif. Distrib. Theory  15(2): 9-22 (2020)

 \bibitem{R2005}
P. Rizomiliotis.
On the design of binary sequences with maximum nonlinear span.
In: Proc. Int'l Symp. Inform. Theory-ISIT2005, 469-473,  IEEE Xplore (2005)


 \bibitem{R2006}
 	P. Rizomiliotis.
Constructing periodic binary sequences with maximum nonlinear span. IEEE Trans. Inform. Theory 52(9): 4257-4261 (2006)




 \bibitem{RK2005}
 P. Rizomiliotis, N. Kalouptsidis:
Results on the nonlinear span of binary sequences. IEEE Trans. Inform. Theory 51(4): 1555-1563 (2005)


\bibitem{SL2000}
C. Seo, S. Lee, Y. Sung, K. Han, S. Kim. A lower bound on the linear span of an FCSR. IEEE Trans. Inform. Theory 46(2): 691-693 (2000)

\bibitem{SW2019}
 	Z. Sun, A. Winterhof.
On the maximum order complexity of the  Thue-Morse and Rudin-Shapiro sequence. Unif. Distrib. Theory 14(2): 33-42 (2019)



\bibitem{SW19b}
Z. Sun, A. Winterhof. On the maximum order complexity of subsequences of the Thue-Morse and Rudin-Shapiro sequence along squares.
 Int. J. Comput. Math. Comput. Syst. Theory  4(1): 30-36  (2019)
		



\bibitem{SZLH2017}
Z. Sun, X. Zeng, C. Li, T. Helleseth. Investigations on periodic sequences
 with maximum nonlinear complexity. IEEE Trans. Inform. Theory 63(10): 6188-6198 (2017)

\bibitem{TQ2010}
T. Tian, W.-F. Qi.
$2$-adic complexity of binary $m$-sequences. IEEE Trans. Inform. Theory 56(1): 450-454 (2010)

\aw{
\bibitem{VCC2022}
M. Vielhaber, M.d.P. Canales Chacón, S. Jara Ceballos. 
Rational complexity of binary sequences, F$\mathbb{Q}$SRs, and pseudo-ultrametric continued fractions in $\mathbb{R}$. Cryptogr. Commun. 14(2): 433-457 (2022)
}

\bibitem{W2010}
A. Winterhof. Linear complexity and related complexity measures.
In: I. Woungang (ed.): Selected Topics in Information and Coding Theory,  3-40, World Scientific, Singapore
(2010)

\bibitem{W2023}
A. Winterhof.
Pseudorandom binary sequences: quality measures and number theoretic constructions. IEICE Trans. Fundam. Electron. Commun. Comput. Sci. DOI:10.1587/transfun.2023SDI0001
(to appear) (2023) 
\aw{
\bibitem{XZLJ2019}
Z. Xiao, X. Zeng, C. Li, Y. Jiang.
Binary sequences with period $N$ and nonlinear complexity $N-2$. 
Cryptogr. Commun. 11(4): 735-757 (2019)
}

\bibitem{XQL2014}
H. Xiong, L. Qu, and C. Li. A new method to compute the $2$-adic complexity of binary sequences. IEEE Trans. Inform. Theory 60(4): 2399-2406 (2014) 






\end{thebibliography}
\end{document}